\documentclass[12pt]{article}
\pdfoutput=1
\usepackage[T1]{fontenc}
\usepackage[latin9]{inputenc}
\usepackage{geometry}
\usepackage{units}
\usepackage{amsmath}
\usepackage{stackrel}
\usepackage{graphicx}
\usepackage{esint}
\usepackage{units}

\begin{document}

\title{Thermodynamic optimization of an electric circuit as a non--steady
energy converter}

\author{Gabriel Valencia--Ortega\textsuperscript{a} and Luis--Antonio Arias--Hernandez\textsuperscript{b}\\
 Departamento de Física, Escuela Superior de Física y Matemáticas,
Instituto\\
 Politécnico Nacional, U. P. Zacatenco, edif. \#9, 2o Piso, Ciudad
de México,\\
 07738, MÉXICO, gvalencia@esfm.ipn.mx\textsuperscript{a} and larias@esfm.ipn.mx\textsuperscript{b}}

\maketitle

\begin{abstract}
Electrical circuits with transient elements can be good examples of
systems where non--steady irreversible processes occur, so in the
same way as a steady state energy converter, we use the formal construction
of the first order irreversible thermodynamic (FOIT) to describe the
energetics of these circuits. In this case, we propose an isothermic
model of two meshes with transient and passive elements, besides containing
two voltage sources (which can be functions of time); this is a non--steady
energy converter model. Through the Kirchhoff equations, we can write
the circuit phenomenological equations. Then, we apply an integral
transformation to linearise the dynamic equations and rewrite them
in algebraic form, but in the frequency space. However, the same symmetry
for steady states appears (cross effects). Thus, we can study the
energetic performance of this converter model by means of two parameters:
the ``force ratio\textquotedblright{} and the ``coupling degree\textquotedblright .
Furthermore, it is possible to obtain the characteristic functions
(dissipation function, power output, efficiency, etc.). They allow
us to establish a simple optimal operation regime of this energy converter.
As an example, we obtain the converter behavior for the maximum efficient
power regime (MPE).

\vspace{0.3cm}
 Keywords: Non--Equilibrium and irreversible thermodynamics; Performance
characteristics of energy conversion systems; figure of merit; Energy
conversion.
\end{abstract}

\section{Introduction}

\begin{figure}
\begin{centering}
\includegraphics{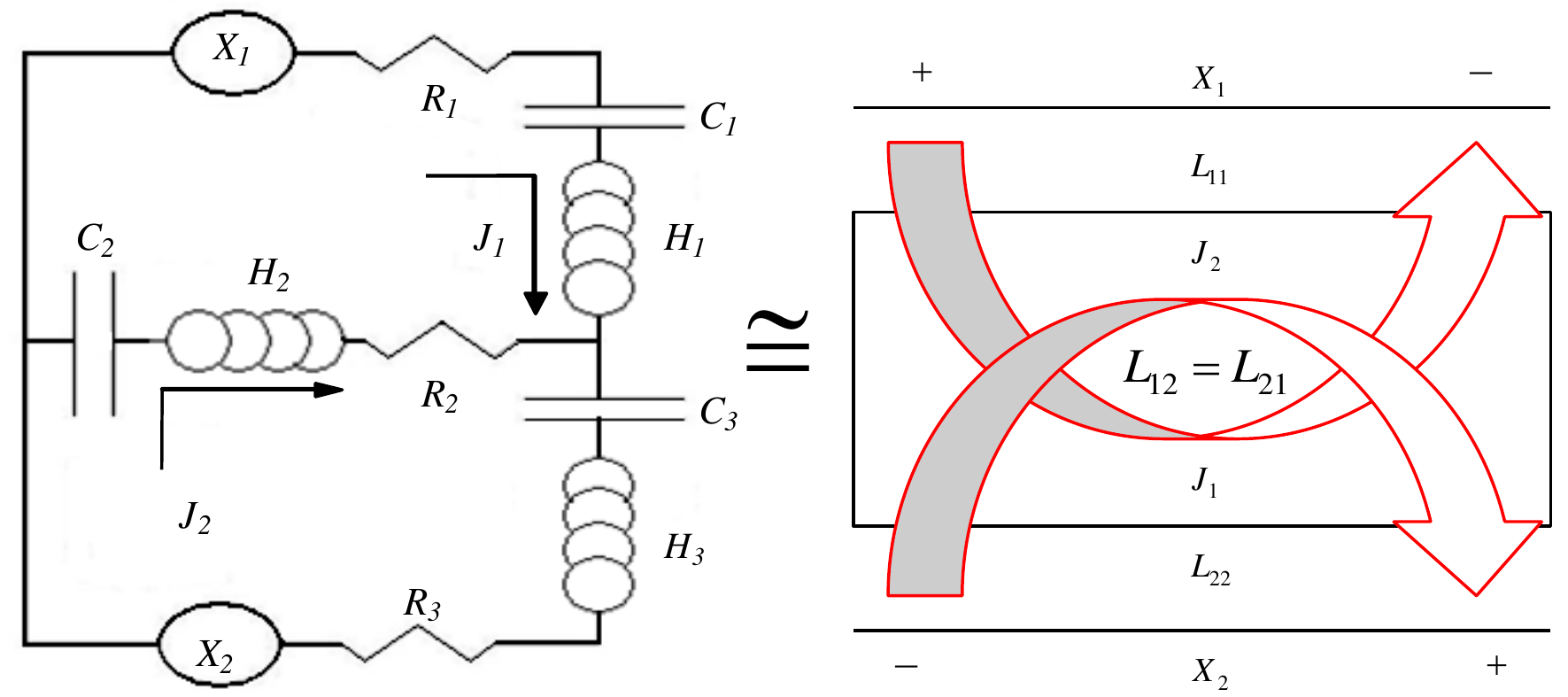} 
\par\end{centering}

\caption{\label{LCRcircuit}LCR--circuit of two meshes as an energy converter
model (left) can be depicted with the scheme on the right side, which
is the simplest energy converter (two fluxes and two forces). Through
the expresion for the entropy production: $\sigma=J_{1}X_{1}+J_{2}X_{2}>0$
and direction of fluxes, we regard that $\left|J_{2}X_{2}\right|>\left|J_{1}X_{1}\right|$
where $X_{1}<0$ is linked with non--spontaneous flux, while $X_{2}>0$
causes the spontaneous flux.}
\end{figure}

The development of Non--Equilibrium Thermodynamics has several objectives,
these objectives go from determining the properties of materials \cite{HernanTovarMejia14,SantamPerezRubi16,GardenGuillouRichardand12}
under different gradients to testing the consistency of the cosmological
models \cite{GonzalezCorderoAngulo16,LandsbergDeVos89,AresLopezAngulo05}.
One of these objetives, which is at the same time a kind of tool,
gives rise to the study of energy conversion. The energy performance
of the converters can be studied based on the entropy production.
Using the form of this quantity we can construct the functions that
characterize its energetic performance \cite{GonzalezAriasAngulo13,Ramadan15}.
In particular, in a linear and steady regime one finds that the energy
performance obeys operating regimes consistent with imposed boundary
conditions; these conditions imply specific relations among its design
and the spontaneous and non--spontaneous gradients on the system \cite{IPrigogine,AriasAnguloPaez08,OdumPinkerton55}.
The optimal points of operation are related with the boundary conditions
and are linked with several optimization criteria (objective functions)
independent from the model. A model with these features, can be used
to describe actual converters \cite{OdumPinkerton55,HoffmannBurzlerSchuber97}.
Nevertheless, it is clear that real processes occur in non--steady
conditions, so it is essential to develop a basic model for these
conditions (see Figure \ref{LCRcircuit}). 

From the Caplan and Essig works \cite{Caplan-Essig}, one can make
a first step toward a description of linear energy converters with
two coupled fluxes and two forces. In general the flows $J_{i}$ displayed
on a real system are usually complicated non--linear functions of
the forces $X_{i}$. However, the linear regime allows us to give
a quite acceptable description of the phenomenon of energy conversion.
The phenomenological equations of this system are given by

\begin{equation}
J_{1}=L_{11}X_{1}+L_{12}X_{2}\label{fluxstat1}
\end{equation}
and

\begin{equation}
J_{2}=L_{21}X_{1}+L_{22}X_{2},\label{fluxstat2}
\end{equation}
where $L_{ij}=\left(\nicefrac{\partial J_{i}}{\partial X_{i}}\right)_{eq}$
are the partial derivatives evaluated at equilibrium, called the phenomenological
coefficients. Furthermore, if it is admitted at this point the Onsager
reciprocal relationship between the phenomenological cross coefficients,
$L_{12}=L_{21}$. Then we can describe the process of energy conversion
of this basic model in the easiest way.

From this linear scheme (Eqs. \ref{fluxstat1} and \ref{fluxstat2}),
Kadem and Caplan have introduced the degree of coupling

\begin{equation}
q^{2}=\frac{L_{12}^{2}}{L_{11}L_{22}}\geq0.\label{couplparam}
\end{equation}

This is a dimensionless paratemer which measures the quality of the
coupling between spontaneous and non--spontaneous fluxes and comes
directly from the thermodynamics' second law (Figure \ref{LCRcircuit})
\cite{AriasAnguloPaez08,Caplan-Essig}. On the other hand, following
the paper published by Stucki \cite{Stucki80}, it is suitable to
take again the definition of the force ratio, given by \cite{Tribus}

\begin{equation}
x=\sqrt{\frac{L_{11}}{L_{22}}}\frac{X_{1}}{X_{2}}=\sqrt{\frac{L_{11}}{L_{22}}}x',\label{forcratio}
\end{equation}
which measures a direct relationship between the two forces through
$x'$, we call $x^{\prime}=\left(\nicefrac{X_{1}}{X_{2}}\right)$
the force quotient. Moreover, in Sec. 2.1 we will be able to write
(\ref{fluxstat1}) and (\ref{fluxstat2}) in terms of the $q$ and
$x$ parameters, although we admit that phenomenological equations
show time dependence.

One of these converters, which can be studied in a regime dependent
on time with a good approximation is an electric circuit with transient
elements, such as capacitors and inductors. The dynamic behavior of
these elements is well--known, and secondly, we know that optimization
criteria do not depend on the model. Then it is possible to build
a non--steady circuit, i.e, one that contains both capacitors and
inductors in order to study its energetic performance. In this paper
(section 2) we have considered an inductive--capacitive--resistive
(LCR) circuit of two meshes, that works through voltage sources which
may even depend on time. In this way the corresponding phenomenological
relationships also depend on time. Due to these equations not being
inverted easily, we should apply a suitable integral transformation,
then they turn algebraic in the frequency space. When we return to
the time space by a reverse transformation, the inverted generalized
equations satisfy the same reciprocal relationships as well as steady
converters \cite{Onsager31}. For this reason, this LCR circuit can
be described using the same model presented in \cite{AriasAnguloPaez08}.
Now we have phenomenological coefficients space time dependent, which
contain information about charging and saturation time of transient
elements in the circuit.

In section 3 we will discuss the energetic behavior of this $(2\times2)$
converter, being that it can work in several operation well--known
regimes (minimum dissipation, maximum power output, maximum efficiency,
maximum ecological function, etc.) \cite{AriasAnguloPaez08,HoffmannBurzlerSchuber97,AnguloAriasSantillan02}.
Additionally, we study the evolution of some characteristic functions
when voltage sources are constant and when they depend periodically
on time, in some of the above regimes. With the same objetive, in
section 4 we focus our attention on another operation regime, which
was derived from the optimization criterion called ``Efficient Power''
proposed by Sucki to study the energetics of the mithochondria \cite{Stucki80}
and used by Yilmaz in the context of Finite Time Thermodynamics (FFT)
\cite{Yilmaz06}. In this section we establish and discuss this mode
of operation for the LCR circuit. Finally, in section 5 we write some
conclusions based in our results.

\section{Energetics of LCR--circuit as a converter with two coupled fluxes}

In this section we estudy the energetics of a $(2\times2)$ system
through the constitutive equations dependent on the time. By means
of the quantities $\left(q,x\right)$ we write their characteristic
functions and shall obtain their temporal evolution. This evolution
is possible because the circuit depends on the nature of the transient
elements and due to the fact $L_{12}(t)=L_{21}(t)$, then we can establish
some optimum modes of operation.

\subsection{Constitutive equations}

Let $J_{1}=J_{1}(t)$ and $J_{2}=J_{2}(t)$ be the same two coupled
generalized fluxes ($J_{1}$ being the driven flux and $J_{2}$ the
driver flux) which appear in the description of steady energy converters;
besides $X_{1}$ and $X_{2}$ are the conjugate generalized forces
associated with the fluxes. For the linear regime, fluxes depend on
the forces by means the Onsager equations. Therewith, if we use the
parameters of design and mode of operation defined by (\ref{couplparam})
and (\ref{forcratio}) we can rewrite (\ref{flux1}) and (\ref{flux2})
as:

\begin{equation}
J_{1}=\sqrt{L_{11}}\left(X_{1}\sqrt{L_{11}}+X_{2}\sqrt{L_{22}}q\right)=\left(\frac{x+q}{x}\right)L_{11}X_{1},\label{flux1}
\end{equation}

\begin{equation}
J_{2}=\sqrt{L_{22}}\left(X_{1}\sqrt{L_{11}}q+X_{2}\sqrt{L_{22}}\right)=\left(1+qx\right)L_{22}X_{2},\label{flux2}
\end{equation}
where $\left(L_{12}=L_{21}\right)$, the Onsager symmetry relation
between crossed coefficients has been used. We note that in the limit
case $q\rightarrow0$, each flux becomes proportional to its proper
conjugate force through the direct phenomenological coefficient, i.e,
cross effects vanish and therefore the fluxes become independent.
And when $q\rightarrow1$, the fluxes tend to a fixed relationship
independently of the magnitudes force \cite{AriasAnguloPaez08,Caplan-Essig}.
Here we have taken $X_{2}>0$ as the associated force to the driver
flux. Whereby, $x$ measures the fraction of $X_{1}<0$ appearing
due to the presence of the flux $J_{2}$. Then, the physical values
for the force ratio parameter are in the interval $\left[-1,0\right]$.
We also can represent $J_{1}$ and $J_{2}$, if we know the force
$X_{2}$, which is carried out in pro the gradient. Henceforth, energy
convertion description will be expressed with this condition.

\subsection{LCR--circuit's dynamic equations}

In the LCR--circuit (see Figure \ref{LCRcircuit}) there are two coupled
fluxes (currents) and two conjugate forces (voltages). The phenomenological
equations of the circuit are given by the Kirchhoff laws; if we apply
the corresponding Kirchhoff equation (energy conservation), for the
mesh 1 (left side) we have:

\begin{equation}
X_{1}=\left(\frac{1}{C_{1}}+\frac{1}{C_{2}}\right)\int J_{1}dt+\left(H_{1}+H_{2}\right)\frac{dJ_{1}}{dt}+\left(R_{1}+R_{2}\right)J_{1}-\frac{1}{C_{2}}\int J_{2}dt-H_{2}\frac{dJ_{2}}{dt}-R_{2}J_{2},\label{Kirch1}
\end{equation}
while for the mesh 2 (right side) the equation is:

\begin{equation}
X_{2}=\left(\frac{1}{C_{2}}+\frac{1}{C_{3}}\right)\int J_{2}dt+\left(H_{2}+H_{3}\right)\frac{dJ_{2}}{dt}+\left(R_{2}+R_{3}\right)J_{2}-\frac{1}{C_{2}}\int J_{1}dt-H_{2}\frac{dJ_{1}}{dt}-R_{2}J_{1}.\label{Kirch2}
\end{equation}

Hereafter, we have considered that the inductors are far enough apart
to neglet the effect of mutual inductances.

\subsubsection{Direct Current LCR system}

First, we consider $X_{1}$ and $X_{2}$ independent on time, then
we note that (\ref{Kirch1}) and (\ref{Kirch2}) are not simple invertible
equations. So, we can use an integral transformation (Laplace transformation)
because it enables direct inclusion of initial conditions for the
circuit. Hence, the above equations are sent to the frequency space
and they become invertible equations:

\begin{equation}
\frac{X_{1}}{s}=\left(\frac{1}{C_{1}}+\frac{1}{C_{2}}\right)\frac{I_{1}(s)}{s}+\frac{q_{0}}{sC_{1}}+s\left(H_{1}+H_{2}\right)I_{1}(s)-H_{1}J(0+)+\left(R_{1}+R_{2}\right)I_{1}(s)-\frac{I_{2}(s)}{sC_{2}}-sH_{2}I_{2}(s)-R_{2}I_{2}(s),\label{kirchfrec1}
\end{equation}
and

\begin{equation}
\frac{X_{2}}{s}=\left(\frac{1}{C_{2}}+\frac{1}{C_{3}}\right)\frac{I_{2}(s)}{s}+\frac{q_{0}}{sC_{3}}+s\left(H_{2}+H_{3}\right)I_{2}(s)-H_{3}J(0+)+\left(R_{2}+R_{3}\right)I_{2}(s)-\frac{I_{1}(s)}{sC_{2}}-sH_{2}I_{1}(s)-R_{2}I_{1}(s),\label{kirchfrec2}
\end{equation}
where $\left(H_{1},\, H_{2},\, H_{3}\right)$, $\left(C_{1},\, C_{2},\, C_{3}\right)$
and $\left(R_{1},\, R_{2},\, R_{3}\right)$ are values of the inductances,
capacitances and resistances in the circuit, respectively; $q_{0}$
is the initial capacitors' charge for each one $C_{1},\, C_{3}$ and
$J(0+)$ is the initial current that energized $H_{1}$ and $H_{3}$.
The system formed by (\ref{kirchfrec1}) and (\ref{kirchfrec2}) can
be inverted, such that the fluxes $I_{1}(s)$ and $I_{2}(s)$ are
expressed in terms of forces $X_{1}$ and $X_{2}$. Therefore the
system is written as follows:

\begin{equation}
\left[\begin{array}{c}
I_{1}(s)\\
I_{2}(s)
\end{array}\right]=\frac{1}{Q(s)}\left[\begin{array}{cc}
H_{b2}\left(s^{2}+\frac{R_{2}+R_{3}}{H_{b2}}s+\frac{1}{H_{b2}}\left(\frac{1}{C_{2}}+\frac{1}{C_{3}}\right)\right) & H_{2}\left(s^{2}+\frac{R_{2}}{H_{2}}s+\frac{1}{C_{2}H_{2}}\right)\\
H_{2}\left(s^{2}+\frac{R_{2}}{H_{2}}s+\frac{1}{C_{2}H_{2}}\right) & H_{b1}\left(s^{2}+\frac{R_{1}+R_{2}}{H_{b1}}s+\frac{1}{H_{b1}}\left(\frac{1}{C_{1}}+\frac{1}{C_{2}}\right)\right)
\end{array}\right]\left[\begin{array}{c}
X_{1}-\frac{q_{0}}{C_{1}}\\
X_{2}-\frac{q_{0}}{C_{3}}
\end{array}\right],\label{matrfreq}
\end{equation}
where $H_{b1}=\left(H_{1}+H_{2}\right)$, $H_{b2}=\left(H_{2}+H_{3}\right)$
and

\[
\begin{array}{c}
Q(s)=H_{1}H_{b2}\left(s^{2}+\frac{R_{1}}{H_{1}}s+\frac{1}{C_{1}H_{1}}\right)\left(s^{2}+\frac{R_{2}+R_{3}}{H_{b2}}s+\frac{1}{H_{b2}}\left(\frac{1}{C_{2}}+\frac{1}{C_{3}}\right)\right)+\\
H_{2}H_{3}\left(s^{2}+\frac{R_{2}}{H_{2}}s+\frac{1}{C_{2}H_{2}}\right)\left(s^{2}+\frac{R_{3}}{H_{3}}s+\frac{1}{C_{3}H_{3}}\right).
\end{array}
\]

\begin{figure}
\begin{centering}
\includegraphics{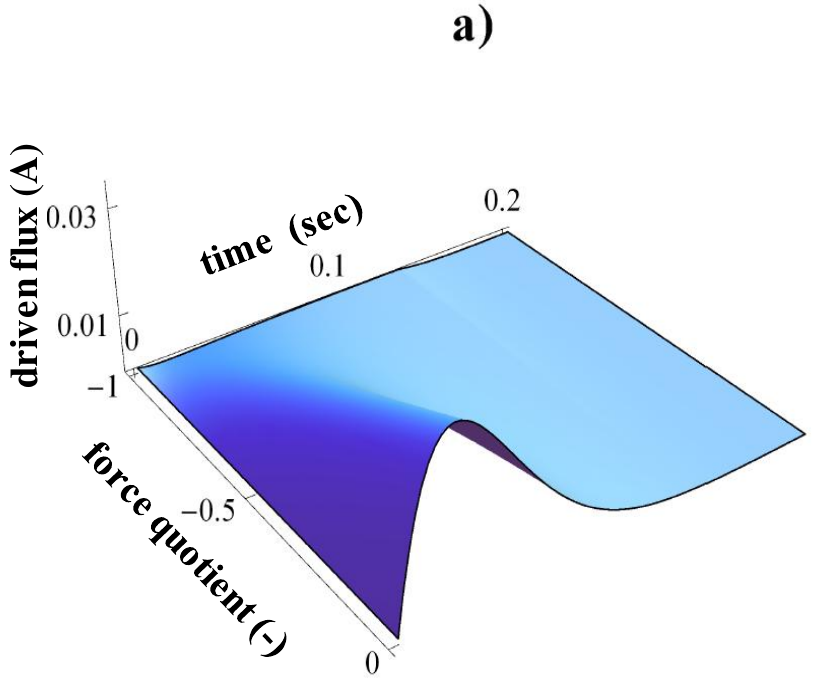}\hspace{3mm}\includegraphics{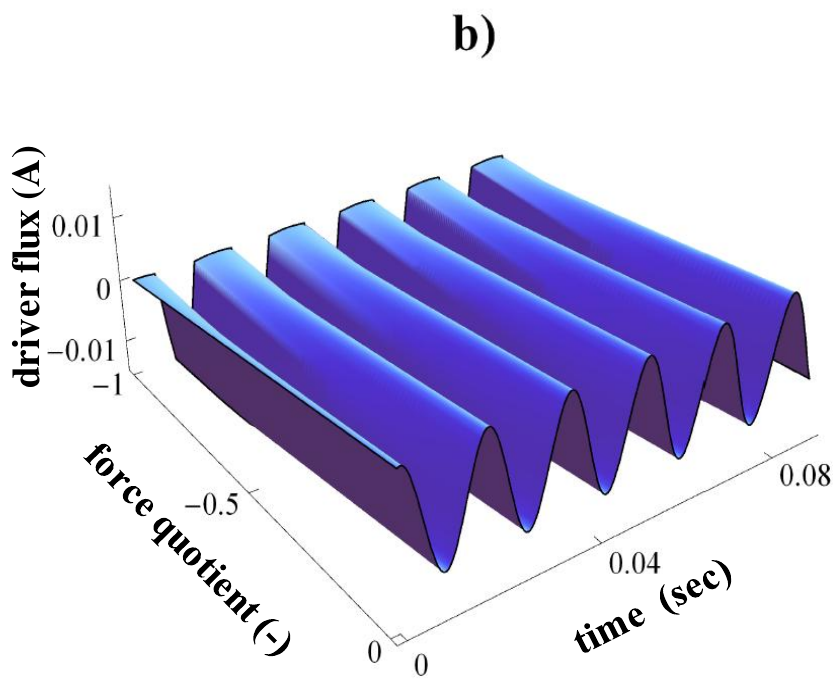} 
\par\end{centering}

\caption{\label{fluxdcacLCR}Time evolution of fluxes a)$J_{1}$ (driven) for
DC--LCR circuit and b) $J_{2}$ (driver) for AC--LCR circuit versus
force quotient $x^{\prime}$. There is a monotonous increasing behavior
of each flux with $x^{\prime}$ in the direct and altern current cases.
While in the temporal evolution of DC case there is a damped behavior
but in AC case this behavior is oscillatory, due to frequency 60Hz
of the sources and the respective values of the circuit elements are:
$H_{1}=0.2\textrm{H},\, H_{2}=4\textrm{H},\, H_{3}=0.1\textrm{H},\, R_{1}=10\Omega,\, R_{2}=680\Omega,\, R_{3}=12\Omega,\,$,
$C_{1}=3\mu F$, $C_{2}=500\mu F$ and $C_{3}=5\mu F$.}
\end{figure}

In addition we have considered that the instant when the voltage sources
are activated the inductors are not energized, that is $J(0+)\rightarrow0$.
Now, in order to obtain the flows $I_{1}(s)$ and $I_{2}(s)$ in the
time space, we use the Heaviside's formula that coincides with the
inverse transform of Laplace and is given by \cite{Edminister}:

\begin{equation}
\mathcal{L}^{-1}\left[\frac{P(s)}{Q(s)}\right]=\stackrel[k=1]{n}{\sum}\frac{P(a_{k})}{Q'(a_{k})}e^{a_{k}t},\label{heavform}
\end{equation}
where every $a_{k}$ are the $n$ different roots of $Q(s)$. We use
this form to express an arbitrary function as polynomial ratios. So,
$P(s)$ and $Q(s)$ are polynomials with the degree of $P(s)$ less
than the degree of $Q(s)$. We use this form of inverse integral transformation
(Eq. \ref{heavform}) on the system formed by (Eq. \ref{matrfreq}).
The new matrix can be represented by the generalized fluxes as follows:

\begin{equation}
\left[\begin{array}{c}
J_{1}\\
J_{2}
\end{array}\right]=\left[\begin{array}{cc}
H_{b2}\overset{4}{\underset{k=1}{\sum}}\frac{\left(a_{k}^{2}+\frac{R_{2}+R_{3}}{H_{b2}}a_{k}+\frac{1}{H_{b2}}\left(\frac{1}{C_{2}}+\frac{1}{C_{3}}\right)\right)e^{a_{k}t}}{Q^{\prime}(a_{k})} & H_{2}\overset{4}{\underset{k=1}{\sum}}\frac{\left(a_{k}^{2}+\frac{R_{2}}{H_{2}}a_{k}+\frac{1}{C_{2}H_{2}}\right)e^{a_{k}t}}{Q^{\prime}(a_{k})}\\
H_{2}\overset{4}{\underset{k=1}{\sum}}\frac{\left(a_{k}^{2}+\frac{R_{2}}{H_{2}}a_{k}+\frac{1}{C_{2}H_{2}}\right)e^{a_{k}t}}{Q^{\prime}(a_{k})} & H_{b1}\overset{4}{\underset{k=1}{\sum}}\frac{\left(a_{k}^{2}+\frac{R_{1}+R_{2}}{H_{b1}}a_{k}+\frac{1}{H_{b1}}\left(\frac{1}{C_{1}}+\frac{1}{C_{2}}\right)\right)e^{a_{k}t}}{Q^{\prime}(a_{k})}
\end{array}\right]\left[\begin{array}{c}
X_{1}-\frac{q_{0}}{C_{1}}\\
X_{2}-\frac{q_{0}}{C_{3}}
\end{array}\right],\label{matronsag}
\end{equation}
with

\begin{equation}
\begin{array}{c}
Q^{\prime}(a_{k})=4\left[H_{1}H_{b2}+H_{2}H_{3}\right]a_{k}^{3}+3\left[R_{1}H_{b2}+H_{1}\left(R_{2}+R_{3}\right)+H_{3}R_{2}+H_{2}R_{3}\right]a_{k}^{2}+\\
2\left[H_{1}\left(\frac{1}{C_{2}}+\frac{1}{C_{3}}\right)+R_{1}\left(R_{2}+R_{3}\right)+\frac{H_{T2}}{C_{1}}+\frac{H_{2}}{C_{3}}+R_{2}R_{3}+\frac{H_{3}}{C_{2}}\right]a_{k}+\\
\left[R_{1}\left(\frac{1}{C_{2}}+\frac{1}{C_{3}}\right)+\frac{\left(R_{2}+R_{3}\right)}{C_{1}}+\frac{R_{2}}{C_{3}}+\frac{R_{3}}{C_{2}}\right]
\end{array}\label{denomRLC}
\end{equation}
and $a_{k}=a_{k}\left(H_{1},\, H_{2},\, H_{3},\, C_{1},\, C_{2},\, C_{3}\, R_{1},\, R_{2},\, R_{3}\right)$.
From (\ref{matronsag}) we note that every matrix element corresponds
to a phenomenological coefficient in FOIT context. For that reason,
$J_{1}$ and $J_{2}$ can be considered as Onsager equations dependent
on time; however $L_{12}(t)=L_{21}(t)$ is satisfied (cross effects).
The system (Eq. \ref{matronsag}) represents the dynamics of the charge
carriers in the circuit when they are stimulated by the potentials
(Figure \ref{fluxdcacLCR}a).

\subsubsection{Altern Current LCR system}

Now we can consider that LCR circuit is fed by alternanting voltage
$X_{1}=X_{1}(t)=X_{01}\textrm{sin}\left(\omega t+\delta\right)$ and
$X_{2}=X_{2}(t)=X_{02}\textrm{sin}\left(\omega t+\beta\right)$ applied
at time $t=0$, besides $\omega$ is the angular frecuency and $\left(\delta,\,\beta\right)$
are the phase angles at time equal to zero, because $\textrm{sin}\left(\omega t+\varphi_{i}\right)$
can express the imaginary component of $\exp\left[i\left(\omega t+\varphi_{i}\right)\right]$.
The system composed by (\ref{Kirch1}) and (\ref{Kirch2}) can be
rewriten as:

\begin{equation}
i\omega X_{01}\exp\left(i\omega t\right)=\left[-\omega^{2}H_{T1}J_{01}+i\omega R_{T1}J_{01}+\left(\frac{1}{C_{1}}+\frac{1}{C_{2}}\right)J_{01}+\omega^{2}H_{2}J_{02}-i\omega R_{2}J_{02}-\frac{J_{02}}{C_{2}}\right]\exp\left(i\omega t\right),\label{KirchacRLC1}
\end{equation}
and

\begin{equation}
i\omega X_{02}\exp\left(i\omega t\right)=\left[-\omega^{2}H_{T2}J_{02}+i\omega R_{T2}J_{02}+\left(\frac{1}{C_{2}}+\frac{1}{C_{3}}\right)J_{02}+\omega^{2}H_{2}J_{01}-i\omega R_{2}J_{01}-\frac{J_{01}}{C_{2}}\right]\exp\left(i\omega t\right),\label{kirchacRLC2}
\end{equation}
where $\delta=\beta=0$, i.e, the voltage sources are not out of phase.
If we divide both equations by $i\omega\exp\left(i\omega t\right)$,
then we obtain a new equations system,

\begin{equation}
\left[\begin{array}{c}
X_{01}\\
X_{02}
\end{array}\right]=\left[\begin{array}{cc}
i\omega H_{T1}+R_{T1}+\frac{1}{i\omega}\left(\frac{1}{C_{1}}+\frac{1}{C_{2}}\right) & -\left(i\omega H_{2}+R_{2}+\frac{1}{i\omega C_{2}}\right)\\
-\left(i\omega H_{2}+R_{2}+\frac{1}{i\omega C_{2}}\right) & i\omega H_{b2}+R_{T2}+\frac{1}{i\omega}\left(\frac{1}{C_{2}}+\frac{1}{C_{3}}\right)
\end{array}\right]\left[\begin{array}{c}
J_{01}\\
J_{02}
\end{array}\right],\label{kirchsysLCR}
\end{equation}
taking into account the circuit impedances: $Z_{12}=Z_{1}+Z_{2}=i\omega H_{b1}+R_{b1}+\frac{1}{i\omega}\left(\frac{1}{C_{1}}+\frac{1}{C_{2}}\right)$,
$Z_{2}=i\omega H_{2}+R_{2}+\frac{1}{i\omega C_{2}}$ and $Z_{23}=Z_{2}+Z_{3}=i\omega H_{b2}+R_{b2}+\frac{1}{i\omega}\left(\frac{1}{C_{2}}+\frac{1}{C_{3}}\right)$,
with $R_{b1}=R_{1}+R_{2}$ and $R_{b2}=R_{2}+R_{3}$ the sum of the
resistances of each mesh. Thus, steady states appear in frecuency
space because current and voltages are linearly related in AC circuits
(Figure \ref{fluxdcacLCR}b).

Therefore, the previous system can be inverted as follows:

\begin{equation}
\left[\begin{array}{c}
J_{01}\\
J_{02}
\end{array}\right]=\left[\begin{array}{cc}
\mid\frac{Z_{2}+Z_{3}}{\triangle}\mid\textrm{sen}\left(\omega t-\theta_{23}\right) & \mid\frac{Z_{2}}{\triangle}\mid\textrm{sen}\left(\omega t-\theta_{2}\right)\\
\mid\frac{Z_{2}}{\triangle}\mid\textrm{sen}\left(\omega t-\theta_{2}\right) & \mid\frac{Z_{1}+Z_{2}}{\triangle}\mid\textrm{sen}\left(\omega t-\theta_{12}\right)
\end{array}\right]\left[\begin{array}{c}
X_{10}\\
X_{20}
\end{array}\right],\label{fluxyfuerzespreal2}
\end{equation}
with $\triangle=\left(Z_{2}+Z_{3}\right)\left(Z_{1}+Z_{2}\right)-Z_{2}^{2}$,
the determinant of (\ref{kirchsysLCR}) and

\[
\theta_{23}=\arctan\left(\frac{\textrm{Im}\left[\frac{Z_{2}+Z_{3}}{\triangle}\right]}{\textrm{Re}\left[\frac{Z_{2}+Z_{3}}{\triangle}\right]}\right),\,\theta_{2}=\arctan\left(\frac{\textrm{Im}\left[\frac{Z_{2}}{\triangle}\right]}{\textrm{Re}\left[\frac{Z_{2}}{\triangle}\right]}\right)\,\textrm{and}\,\theta_{12}=\arctan\left(\frac{\textrm{Im}\left[\frac{Z_{1}+Z_{2}}{\triangle}\right]}{\textrm{Re}\left[\frac{Z_{1}+Z_{2}}{\triangle}\right]}\right).
\]
From (\ref{fluxyfuerzespreal2}) we remark again the time dependence
of the coefficients $L_{ij}$.

\section{Energetic of the LCR--circuit}

Now we take this linear isothermic--isobaric model of conversion process
and build its characteristic functions in terms of $q(t)$ and $x(t)$
for determining the performance condition corresponding to the maximization
of a certain objective function. Firstly, we take into account the
temporal evolution of the above mentioned parameters. From (\ref{matronsag})
and (\ref{fluxyfuerzespreal2}), we study the behavior when the inductors
are saturated, so the circuit turns purely dissipative. On the other
hand, when the capacitors are saturated, there are no longer currents.
The saturation time is related to the circuit characteristic time.
For the study of the energetics of the converter we consider characteristic
functions as the dissipation function, the power output and efficiency,
as well as objective functions as the ecological function and the
efficient power. These functions can be written in terms of parameters
$\left[q(t),\, x(t)\right]$.

We have considered the conventional values of inductances, capacitances
and resistances reported in literature of electric circuits \cite{Edminister};
in this case we take the following values: $H_{1}=0.2\textrm{H},\, H_{2}=4\textrm{H},\, H_{3}=0.1\textrm{H},\, R_{1}=10\Omega,\, R_{2}=680\Omega,\, R_{3}=12\Omega,\,$,
$C_{1}=3\mu F$, $C_{2}=500\mu F$ and $C_{3}=5\mu F$.

\subsection{Efficiency}

We define the efficiency of the thermodynamic process as the ratio
of useful energy output divided by energy input, which is in general
supplied to the system. If we take the Caplan and Essig idea \cite{Caplan-Essig},
efficiency can be obtained in terms of Onsager relations as follows:

\begin{equation}
\eta[L_{ij}(t)]=-\frac{TJ_{1}X_{1}}{TJ_{2}X_{2}}=-\frac{x(t)\left[x(t)+q(t)\right]}{q(t)x(t)+1}=-\frac{\left(x'\right)^{2}L_{11}(t)+x'L_{12}(t)}{x'L_{12}(t)+L_{22}(t)}.\label{effic}
\end{equation}

\begin{figure}
\begin{centering}
\includegraphics{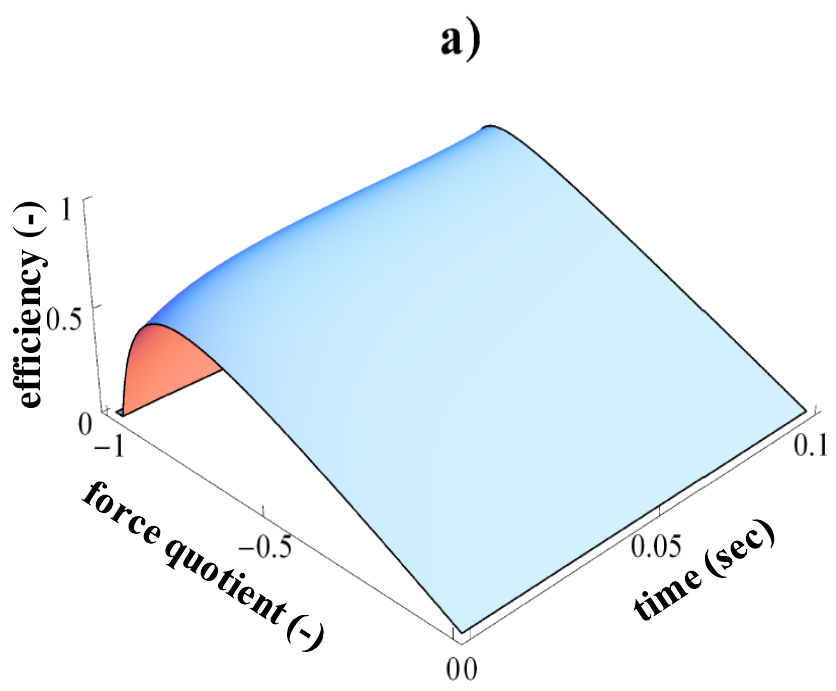}\hspace{3mm}\includegraphics{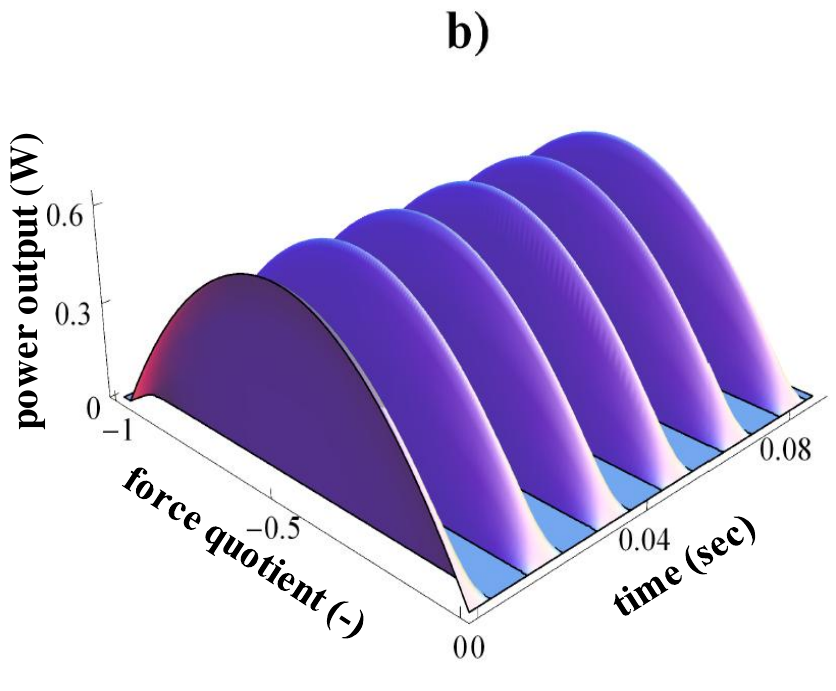} 
\par\end{centering}

\caption{\label{efficdcypotsalacLCR}a) Efficiency for DC--LCR circuit and
b) power output for AC--LCR circuit for $X_{2}^{2}$ fixed and $T=300\textrm{K}$
versus time and $x'$.}
\end{figure}

The graph (Figure \ref{efficdcypotsalacLCR}a) of this function is
a convex surface with only one maximum point. This point shows an
exclusive relationship between the input and output fluxes which maximize
the efficiency for some Onsager coefficients. On the other hand, the
temporal behavior is monotonically increasing and tends to an asymptotic
value. In the AC circuit, efficiency has a maximum for the same value
of $x^{\prime}$, although it becomes a periodic function of time
(not shown here).

\subsection{Power output}

For an isothermal process of two coupled fluxes this energetic function
is given by \cite{OdumPinkerton55,SantillanAriasAngulo99}

\begin{equation}
P[L_{ij}(t)]=-TJ_{1}X_{1}=-x(t)\left[x(t)+q(t)\right]TL_{22}(t)X_{2}^{2}=-\left[\left(x'\right)^{2}L_{11}(t)+x'L_{12}(t)\right]TX_{2}^{2}.\label{powout}
\end{equation}

This equation corresponds to a convex surface with a maximum absolute
value at $x'$, where it is clearly shown that this maximum is maintained
with the same value along Figure\ref{efficdcypotsalacLCR}b, i.e,
the circuit has reached a steady state, because of the circuit's characteristic
time is less than the time corresponding to $60\,\textrm{Hz}$ . In
the DC circuit (which is not shown here) this function reaches a maximum
value at the same $x'$. However, there is a damped oscillation in
time corresponding to another aspect of the same phenomenon. This
means, that the energetic compromise of the system is only due to
the parameter $x'$.

\subsection{Dissipation function}

The entropy production for a $(2\times2)$ system of fluxes and forces
in general is given by \cite{Caplan-Essig,DeGrootMazur},

\begin{equation}
\sigma=J_{1}X_{1}+J_{2}X_{2}\label{dissfunc}
\end{equation}
and as mentioned by several authors \cite{Caplan-Essig,Stucki80},
the dissipation function for isothermal systems is $\Phi=T\sigma$.
By the substitution of (\ref{flux1}) and (\ref{flux2}) into (\ref{dissfunc})
we obtain $\Phi$ in terms of the parameters $q(t)$, $x(t)$ and
the coefficients $L_{ij}(t)$, namely

\begin{equation}
\Phi[L_{ij}(t)]=\left[x^{2}(t)+2x(t)q(t)+1\right]TL_{22}(t)X_{2}^{2}=\left[\left(x'\right)^{2}L_{11}(t)+2x'L_{12}(t)+L_{22}(t)\right]TX_{2}^{2}.\label{prodentro}
\end{equation}

\begin{figure}
\begin{centering}
\includegraphics{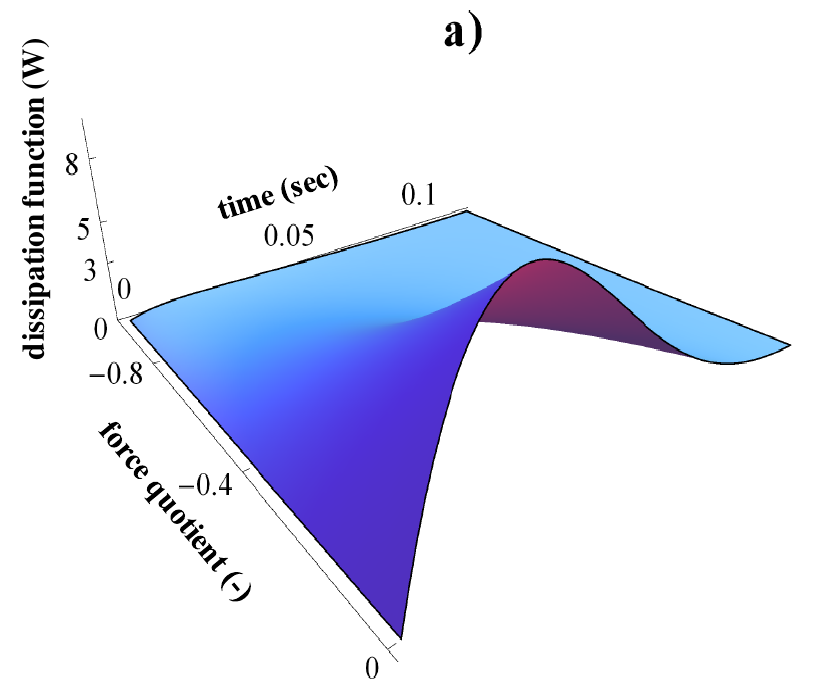}\hspace{3mm}\includegraphics{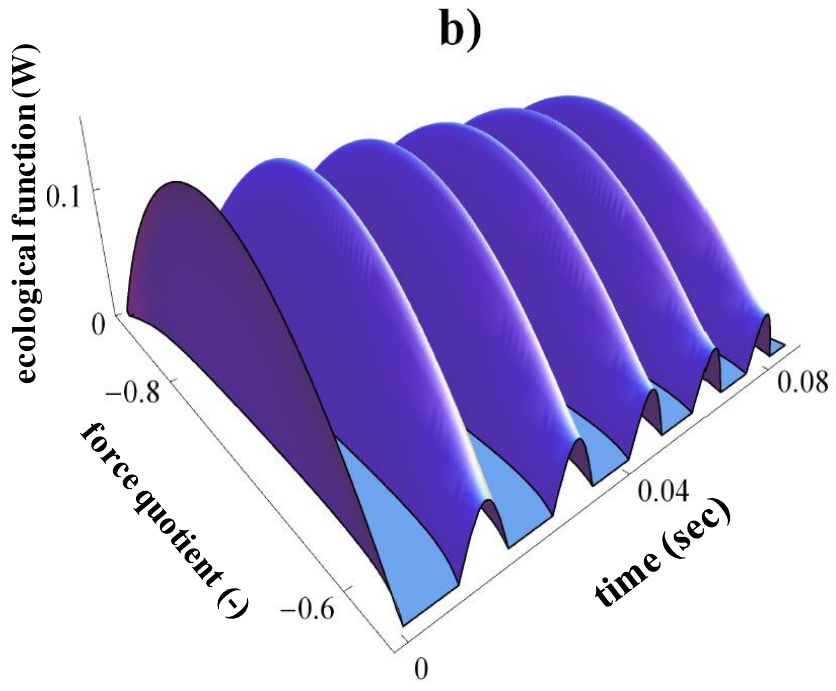} 
\par\end{centering}

\caption{\label{dissifuncdcyecolfunacLCR}a) Dissipation function for DC--LCR
circuit and b) ecological function AC--LCR circuit for $X_{2}^{2}$
fixed and $T=300\textrm{K}$ versus time and $x'$. We observe that
the evolution over time is determined by the dynamics of each system.}
\end{figure}

This function is a concave surface with a minimum value with respect
to force quotient $x'$ as well as the steady state too (see Figure
\ref{dissifuncdcyecolfunacLCR}a). In the AC case, the same value
is obtained but it appears with a frequency of 60 times per second.
This surface is not shown here.

\subsection{Ecological function}

Defining the ecological function as an objetive function ($E=P-\Phi$)
\cite{Angulo91}, by means of (\ref{prodentro}) and (\ref{powout})
we get

\begin{equation}
E[L_{ij}(t)]=-\left[2x^{2}(t)+3x(t)q(t)+1\right]TL_{22}(t)X_{2}^{2}=-\left[\left(x'\right)^{2}L_{11}(t)+3x'L_{12}(t)+L_{22}(t)\right]TX_{2}^{2}.\label{ecolfunct}
\end{equation}

This equation also corresponds to a convex surface with only a maximum
point at $x'$. Now, we see in Figure \ref{dissifuncdcyecolfunacLCR}b
as well as for the similar direct current case (damped oscillation,
which is not shown here), that a good compromise between power output
and dissipated energy is modulated by the presence of transient elements
and the oscillatory behavior over time is kept.

\subsection{Efficient power output}

Another objective function to treat is the efficient power, defined
as the direct product between power output and efficiency \cite{Stucki80,Yilmaz06},
using (\ref{powout}) and (\ref{effic}) we get:

\begin{equation}
P_{\eta}[L_{ij}(t)]=\frac{x^{2}(t)\left[x(t)+q(t)\right]^{2}TL_{22}(t)X_{2}^{2}}{q(t)x(t)+1}=\frac{\left[\left(x'\right)^{2}L_{11}(t)+x'L_{12}(t)\right]^{2}TX_{2}^{2}}{x'L_{12}(t)+L_{22}(t)}.\label{efficpow}
\end{equation}

\begin{figure}
\begin{centering}
\includegraphics{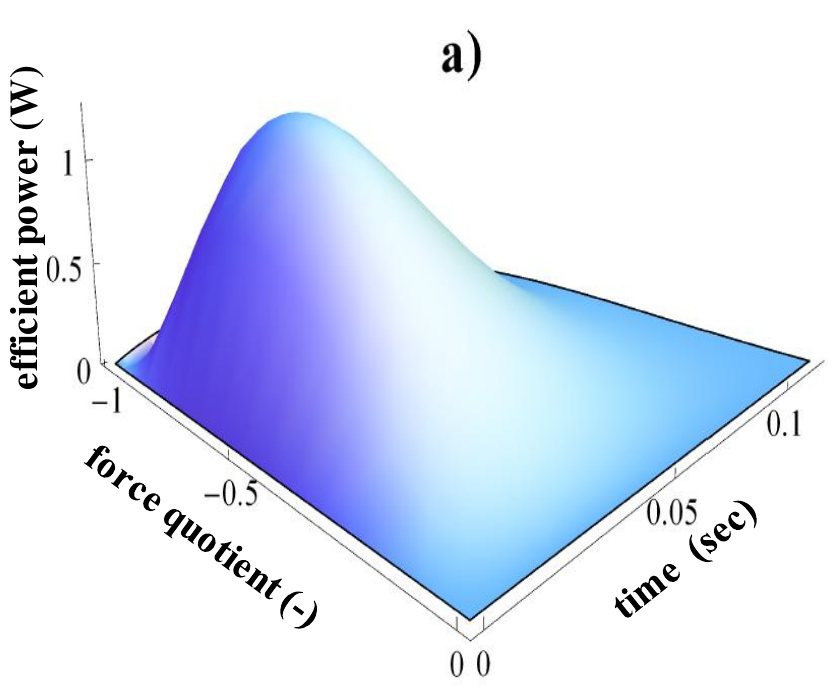}\hspace{3mm}\includegraphics{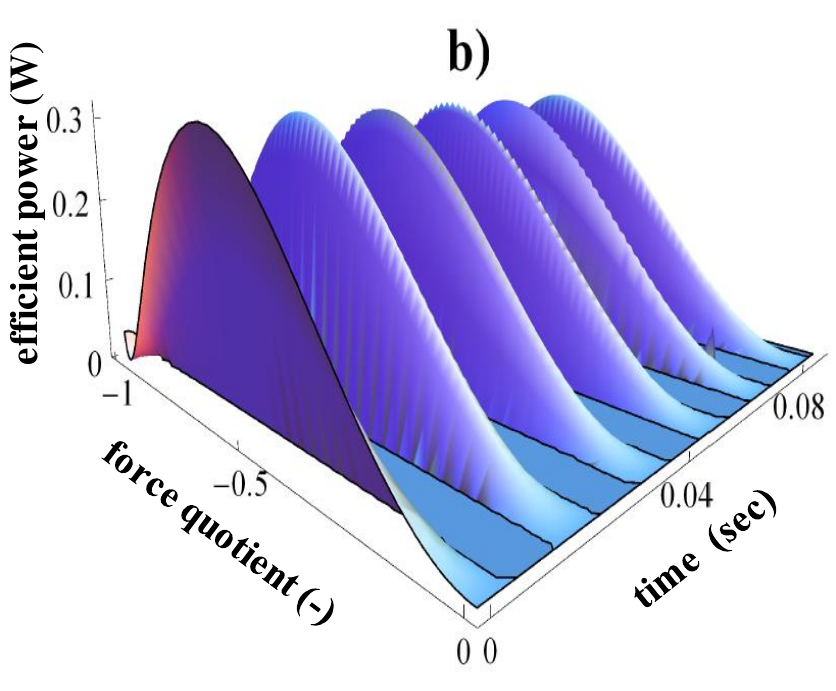} 
\par\end{centering}

\caption{\label{efficpowdcyacLCR}Efficient power for a) DC and b) AC LCR circuit
for $X_{2}^{2}$ fixed and $T=300\textrm{K}$ versus time and $x'$.
Efficient power reaches a maximum value at an especific $x'$, means,
there is a unique fraction of $X_{1}$ due to the presence of $X_{2}$
in order to obtain a maximum efficient power regime.}
\end{figure}

In Figure \ref{efficpowdcyacLCR} we observe that the surfaces for
power output and efficiency have a maximum point. Then the efficient
power presents a convex behavior with a maximum value too (see Figure
\ref{efficpowdcyacLCR}). This regime will be analyzed in the next
Section. Nevertheless, one can see the same behavior with respect
to the parameter $x^{\prime}$, which is related to the exchange of
energy, as the temporal behavior is governed by the nature of the
sources.

From the values for inductances, capacitances and resistances set
out at the begining of this section, we verify in Figures \ref{efficdcypotsalacLCR},
\ref{dissifuncdcyecolfunacLCR}, and \ref{efficpowdcyacLCR} the well--know
temporal behavior in circuit theory, furthermore those functions agree
with the characteristic curves for steady states. The values of the
circuit elements were taken in order to obtain a coupling between
the circuit meshes.

Finally, when we take $C_{i}\rightarrow\infty$, that is, any sources
or sinks of potential differences appear, we can recover the results
for a LR circuit. While $H_{i}=0$, there are no source or sink of
currents in the circuit, the LCR circuit becomes a CR circuit.

\section{A performance mode of the LCR converter}

From the figures in the previous section, we see that every characteristic
function reaches maximum or minimal values, which are found by means
of $\nicefrac{dP_{\eta}(t)}{dx}\mid_{x_{MEP}}=0$. In this case the
force ratio value that maximizes the efficient power is given by

\begin{equation}
x_{MEP}(t)=-\frac{4+q^{2}(t)-\sqrt{16-16q^{2}(t)+q^{4}(t)}}{6q(t)}.\label{forratmpe}
\end{equation}

\begin{figure}
\begin{centering}
\includegraphics{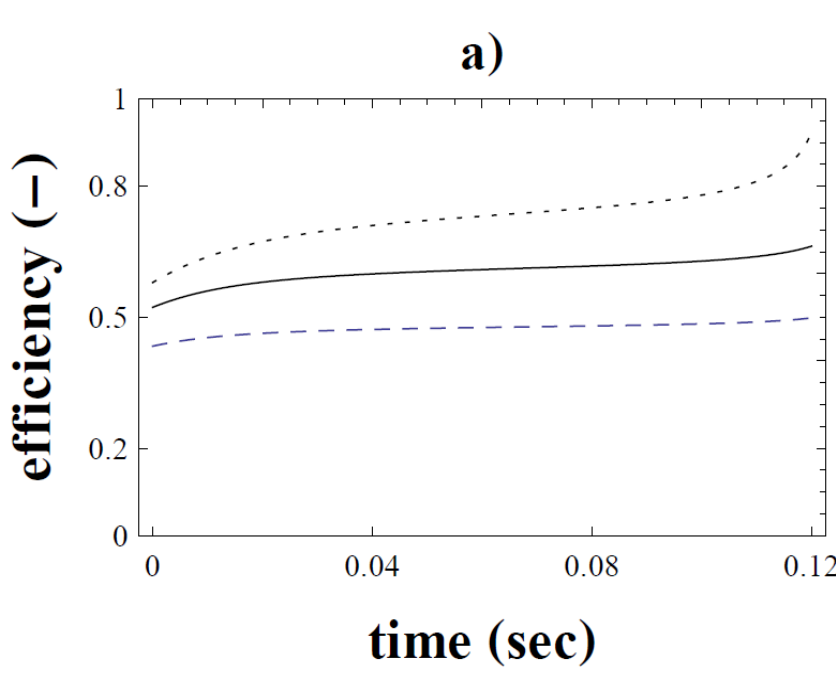}\hspace{3mm}\includegraphics{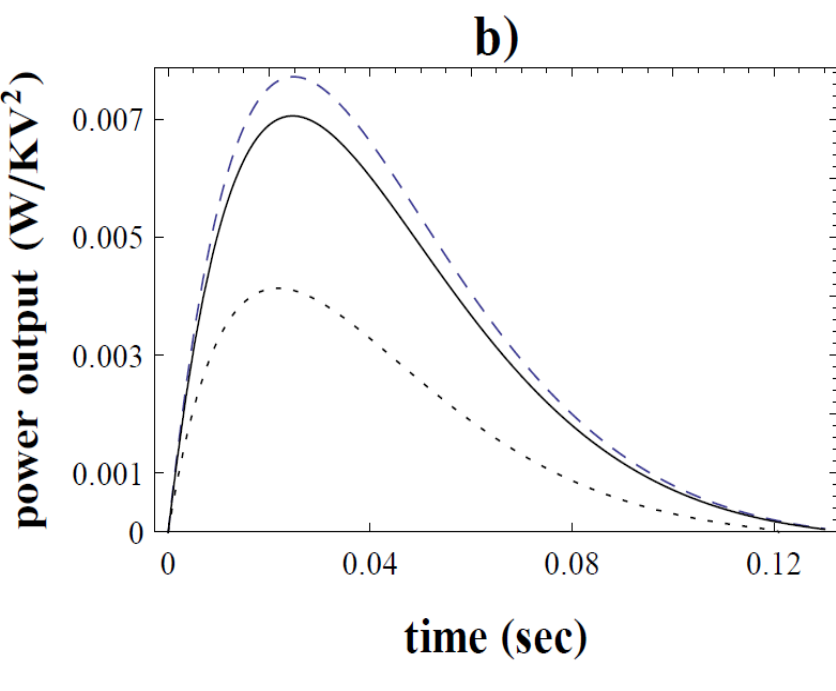} 
\par\end{centering}

\caption{\label{energfuncmep}Normalized energetic functions (efficiency and
power output) are evaluated in the regimes: OE $\left(x_{OE}=-\nicefrac{q(t)}{1+\sqrt{1-q^{2}(t)}}\right)$
(dotted line), MPO $\left(x_{MPO}=-\nicefrac{q(t)}{2}\right)$ (dashed
line) and MEP (solid line) for $TX_{2}^{2}$ fixed versus time, for
the DC case. Power output evaluated in each regimen reaches a maximum
point at different time values.}
\end{figure}

\begin{figure}
\begin{centering}
\includegraphics{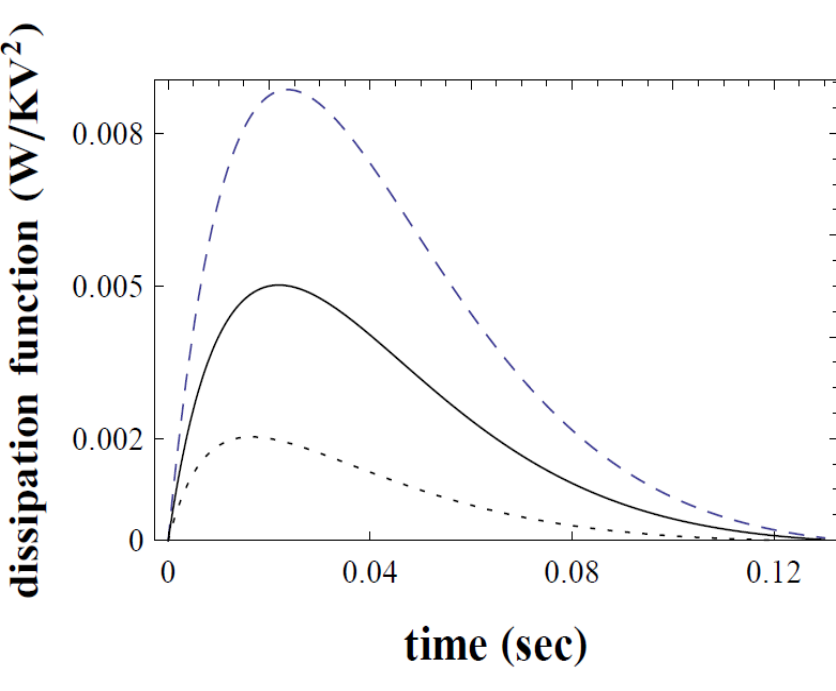} 
\par\end{centering}

\caption{\label{dissipfuncmep}Normalized dissipation function is evaluated
in the regime: oe $\left(x_{OE}=-\nicefrac{q(t)}{1+\sqrt{1-q^{2}(t)}}\right)$(dotted
line), MPO $\left(x_{MPO}=-\nicefrac{q(t)}{2}\right)$ (dashed line)
and MEP (solid line) for $TX_{2}^{2}$ fixed versus time, for the
DC case. We see that dissipation function evaluated in each regimen
reaches a maximum point at different time values.}
\end{figure}

Replacing this force ratio value in each of the expressions for the
characteristic functions, one can ensure that the converter operates
in the maximum efficient power regime (MEP) \cite{Valencia}, hence:

\begin{equation}
\eta_{MEP}(t)=-\frac{4\left(2+q^{2}(t)+\alpha(t)\right)}{3q^{2}(t)},\label{efficmep}
\end{equation}

\begin{equation}
P_{MEP}(t)=\frac{\left(4+q^{2}(t)-\alpha(t)\right)\left(-4+5q^{2}(t)+\alpha(t)\right)}{36q^{2}(t)}TL_{22}(t)X_{2}^{2},\label{potoutmep}
\end{equation}

\begin{equation}
\Phi_{MEP}(t)=\frac{\left[-5+\left(-4+\alpha(t)\right)q^{4}(t)+\left(-2+\alpha(t)\right)5q^{2}(t)\right]}{18q^{2}(t)}TL_{22}(t)X_{2}^{2},\label{funcdissmep}
\end{equation}

\begin{equation}
E_{MEP}(t)=\frac{\left[7q^{4}(t)+\left(26-7\alpha(t)\right)q^{2}(t)+8\left(-4+\alpha(t)\right)\right]}{18q^{2}(t)}TL_{22}(t)X_{2}^{2},\label{ecolfuncmep}
\end{equation}

\begin{equation}
P_{\eta MPE}(t)=\frac{\left(4+q^{2}(t)-\alpha(t)\right)^{2}\left(-4+5q^{2}(t)+\alpha(t)\right)^{2}}{\left(-2+q^{2}(t)-\alpha(t)\right)216q^{4}(t)}TL_{22}(t)X_{2}^{2},\label{efficpotmep}
\end{equation}
where $\alpha(t)=\sqrt{16-16q^{2}(t)+q^{4}(t)}$. In the graphs of
Figures \ref{energfuncmep} and \ref{dissipfuncmep}, 

we note that process variables of LCR--circuit evaluated in the regime
of operation: maximum power output (MPO), optimum efficiency (OE)
and MPE are in function of the circuit characteristic time. Moreover
power output and dissipation function evaluated in each regime are
governed by a damping behavior in the circuit. For that reason, the
energy conversion process stops at some multiple of the circuit characteristic
time. In Figure \ref{energfuncmep}a we show that $\eta_{MPO}<\eta_{MEP}<\eta_{OE}$.
This means that MPE regime can offer an attainable optimum compromise
between energy output and energy input, even for non--stationary energy
converters. Otherwise, in Figs. \ref{energfuncmep}b and \ref{dissipfuncmep}

the MPE regime displays a good coupling between the driver flux and
the driven flux, in order to obtain the power output. In the case
of AC, all the functions keep the same shape but time evolution is
totally oscillatory.

\section{Conclusions.}

In this work we have studied an energy converter with non--steady
processes. This is due to transient elements in the converter (LCR--circuit).
Therefore, the energy conversion depends on time. Nevertheless, there
is a time longer than the characteristic time of the circuit for which
this converter reaches a steady state. In the first place, linearization
of the circuit's dynamic equations (Kirchhoff equations) allow us
to construct generalized flows and forces. The Onsager symmetry relation
between cross coefficients ($L_{12}(t)=L_{21}(t)$) are found at any
time. The evolution of the converter is governed by the behavior of
the coupling coefficients. When the inductors are saturated, $q$
asymptotically reaches a constant value in time and the energy exchange
is fixed. On the other hand when capacitors are saturated, $q$ decays
until the energy exchange stops. Efficient power as a performance
mode of an energy converter could be a suitable regime of operation,
because efficiency evaluated in MPE regime is bigger than MPO regime,
with a small decrement in power output and a large decrement in dissipated
energy.

From these results we will be able to study other converters of the
same nature with an appropriate energetic performance. The damped
behavior of the circuit is due to the values taken for the passive
and transients elements, because we must ensure that this converter
operates in a coupled regime.

\section*{Acknowledgment}

This work was supported by SIP, COFAA, EDI--IPN--MÉXICO and SNI--CONACyT--MÉXICO.

\end{document}